\documentclass[fp]{jpsj3}
\usepackage{txfonts}

\title{Roles of Interfering Radiation Emitted from Decaying Pulses Obeying Soliton Equations Belonging to the 
Ablowitz-Kaup-Newell-Segur Systems}

\author{Hironobu Fujishima$^1$\thanks{fujishima@lamp.is.utsunomiya-u.ac.jp}, Tetsu Yajima$^2$\thanks{yajimat@is.utsunomiya-u.ac.jp}}
\inst{$^1$Optics R\&D Center, CANON INC., 23-10 Kiyohara Kogyoudanchi,
Utsunomiya, Tochigi 323-3298, Japan \\
$^2$Department of Information Systems Science, Graduate School of
Engineering, Utsunomiya University, Yoto 7-1-2, Utsunomiya, 
Tochigi 321-8585, Japan} 

\abst{The nonlinear Schr\"odinger (NLS) equation under the box-type initial condition, which models general multiple pulses deviating from pure solitons, is analyzed. Following the approximation by splitting the initial pulse into many small bins [G. Boffetta and A. R. Osborne, J. Comp. Phys. {\bf 102}, 25 (1992)], we can analyze the Zakharov-Shabat eigenvalue problem to construct transfer matrices connecting the Jost functions in each interval without direct numerical computation. We can obtain analytical expressions for the scattering data that describe interfering radiation emitted from initial pulses. The number of solitons that appear in the final stage is predicted theoretically, and the condition generating an unusual wave such as a double-pole soliton is derived. Numerical analyses under box-type initial conditions are also performed to show that the interplay between the tails from decaying pulses can affect the asymptotic profile.}

\begin{document}
\maketitle

\section{Introduction}
The theory of solitons has played a prominent role in the development of mathematical physics\cite
{ref:Ablowitz1,ref:Calogero,ref:Novikov,ref:Newell,ref:Faddeev,ref:Clarkson}. It has been applied to many 
interesting fields of physics, ranging from high-energy and gravitational physics\cite
{ref:Rajaraman,ref:Manton,ref:Shifman,ref:Dunajski,ref:Weinberg,ref:Belinski} to the physics of more experimentally 
accessible energy scales such as fluid mechanics and plasma physics\cite
{ref:Whitham,ref:Karpman,ref:Lamb,ref:Infeld,ref:Dauxois,ref:Ablowitz2}. In the context of condensed matter or low-temperature physics, Bose-Einstein condensed (BEC) systems are of particular interest and have attracted 
considerable attention. In BEC systems, the macroscopic wave functions of the condensates are known to obey the nonlinear Schr\"odinger equation (NLSE) of the third order, whose nonlinear term represents the biparticle collision of 
constituent atoms\cite{ref:Griffin,ref:Pitaevskii,ref:Pethick,ref:Ueda}. The absolute square of the macroscopic 
wave function is interpreted as the particle density of the constituent atoms and is directly observable by an optical method. Furthermore, by developing many techniques to control the system, experimentalists have already 
realized various geometries by applying ingenious external potentials to the BEC systems. Quasi-one-dimensional potentials 
are among them, and self-focusing bright solitons have been realized by a number of laboratories\cite
{ref:Khaykovich,ref:Strecker,ref:Marchant,ref:Medley,ref:Khawaja,ref:Cornish}. Recently, collisions between such 
BEC solitons have been experimentally examined and reported\cite{ref:Nguyen,ref:Parker,ref:Billam}. Similar 
experiments are also promising possible in the field of nonlinear optics\cite
{ref:Hasegawa,ref:Abdullaev,ref:Agrawal,ref:Kivshar,ref:Taylor}.

In real experiments, however, one cannot generate pure solitons that are the exact solutions of the NLSE, and the 
word ``solitons'' is often used to mean multiple pulses of condensates or photons. Realistically, 
one expects that the multiple pulses that deviate from pure solitons decay, emitting radiation, and  fall into a final state that includes pure solitons after some transition time. The investigation of the roles played by 
the interaction between the emitted radiation is the motivation of this study. For example, even if each 
constituent pulse is too small to sustain a soliton, the interfering radiation and some nonlinear effects can generate a large amplitude. It is expected to observe the creation of new solitons. 

The time evolution of such nonlinear systems is investigated by solving the initial value problems of 
corresponding soliton equations. The inverse scattering transform (IST) is a useful method that can deal with such 
problems\cite{ref:Ablowitz1,ref:Calogero,ref:Novikov,ref:Newell,ref:Faddeev,ref:Clarkson}. This method is based on 
a scattering problem of a set of auxiliary linear equations, which are associated with the original soliton 
equation:
\begin{subequations}
\label{eq:laxeq}
\begin{align}
\label{eq:laxeq_space}
\Psi_x&=M\Psi,\\
\label{eq:laxeq_time}
\Psi_t&=N\Psi,
\end{align}
\end{subequations}
where the quantities $M$ and $N$ are matrices or operators including the unknown functions of the soliton equation and 
the spectral parameter. The wave function $\Psi$ represents an auxiliary field obeying appropriate boundary 
conditions. An important step of the IST method is to analyze the spatial equation (\ref{eq:laxeq_space}) as a 
scattering problem, whose potential term is given by the initial condition of the unknown function. The wave 
function $\Psi$ and the spectral parameter correspond to the eigenfunction and eigenvalue, respectively. This is called 
the Zakharov-Shabat (ZS) problem\cite{ref:Zakharov}. Once the ZS problem is solved, the time-evolved wave function 
is easily obtained through Eq.~(\ref{eq:laxeq_time}), and  the solution of the Cauchy problem is provided by virtue 
of the Gel'fand-Levitan-Malchenko (GLM) equation\cite{ref:Gel'fand}. The GLM equation clearly shows that the number 
of discrete eigenvalues determines that of solitons to be generated asymptotically.

Usually, the ZS problems accompanied by general initial conditions are difficult to solve, and it is rarely 
possible for us to predict how many solitons remain in the final state, except under pure soliton initial 
conditions. Concerning this problem, in 1991, Boffetta and Osborne proposed \cite{ref:Boffetta} an approximation method to obtain the scattering data for arbitrary initial wave packets by discretizing the spatial 
coordinate. Obviously, this method is applicable to other Ablowitz-Kaup-Newell-Segur (AKNS) soliton equations
\cite{ref:AKNS}. In this paper, we consider a set of box-type potentials as models for multiple pulses that are 
not pure solitons by applying this method. We can obtain an analytical expression that describes interfering radiation emitted from decaying original pulses. In addition, we can extract information such as the number of solitons that appear in the final state. We 
find that the interplay between the diffusing tails from decaying pulses can affect the asymptotic profile 
markedly, and this is confirmed by directly integrating the NLSE by numerical simulation. Furthermore, we derive 
the parameter conditions that generate double-pole solitons\cite{ref:Olmedilla}.

This paper is organized as follows. In the next section, we briefly summarize the IST method and the ZS 
problem taking the NLS equation as an example. In Sect.~3, we introduce Boffetta and Osborne's method and 
explain how to extract approximated scattering data. In Sect.~4, distributions of eigenvalues for the 
NLSE with double-box-type initial conditions are derived. We also show that this simple application leads to some nontrivial results including conditions for generating double-pole solitons and the crucial roles played by interfering 
radiation from each boxlike pulse. The results of our numerical simulation are shown in Sect.~5. The final section is 
devoted to discussion and concluding remarks.
\section{Summary of the IST Method and ZS Problem}
We give a brief explanation of the IST method and ZS problem for later use. The theory treated here is no more than textbook knowledge\cite{ref:Ablowitz1,ref:Calogero,ref:Novikov,ref:Newell,ref:Faddeev,ref:Clarkson}. Throughout this paper, we 
take the NLS equation as an illustration:
\begin{equation}
i\psi_t=-\psi_{xx}-2|\psi|^2\psi.\label{NLSE}
\end{equation}
For the NLS equation, the matrices $M$ and $N$ in Eq.~(\ref{eq:laxeq}) are given as
\begin{subequations}
\label{eq:nlslax_matrices}
\begin{align}
\label{eq:nlslax_space}
M&=
\begin{bmatrix}
-i\xi&i\psi^*\\
i\psi&i\xi
\end{bmatrix},\\
\label{eq:nlslax_time}
N&=
\begin{bmatrix}
2i\xi^2-i|\psi|^2&\psi^*_x-2i\xi\psi^*\\
-\psi_x-2i\xi\psi&-2i\xi^2+i|\psi|^2
\end{bmatrix},
\end{align}
\end{subequations}
where $\xi$ is the spectral parameter. Equations (\ref{eq:laxeq_space}) and (\ref{eq:nlslax_space}) completely 
define the ZS problem for the NLS equation. Satsuma and Yajima extensively studied this problem and obtained exact results for $\mathrm{Asech}$-type initial conditions, where $\mathrm{A}$ is a magnitude of the initial pulse\cite{ref:Satsuma}. Other soliton equations belonging to the AKNS system have similar ZS 
problems. 
\par
We introduce the usual boundary condition for $\psi$:
\begin{equation}
 \psi\to0,\quad\mbox{as $|x|\to\infty$.}
\end{equation}
With this boundary condition, each element of the wave function $\Psi$ must become a plane wave. As the fundamental 
solutions, we can select two sets of functions ${\{\phi,\bar{\phi}\}}$ and ${\{\chi,\bar{\chi}\}}$ called the Jost 
functions, which satisfy the boundary conditions
\begin{subequations} 
\label{eq:jost_boundaries}
\begin{align}
\label{eq:jost_neginftyx}
 &\phi(x;\xi)\to\begin{bmatrix}e^{-i\xi x}\\0\end{bmatrix},\ 
 \bar\phi(x;\xi)\to\begin{bmatrix}0\\e^{i\xi x}\end{bmatrix},
 &\mbox{as $x\to-\infty$},\\
\label{eq:jost_posinftyx}
 &\chi(x;\xi)\to\begin{bmatrix}0\\e^{i\xi x}\end{bmatrix},\ 
 \bar\chi(x;\xi)\to\begin{bmatrix}e^{-i\xi x}\\0\end{bmatrix},
 &\mbox{as $x\to+\infty$}.
\end{align}
\end{subequations}
The Jost functions are related to each other as
\begin{equation}\label{eq:scattering_rels}
\begin{split}
\phi(x;\xi)=a(\xi)\bar\chi(x;\xi)+b(\xi)\chi(x;\xi),\\
\bar\phi(x;\xi)=\bar a(\xi)\chi(x;\xi)-\bar b(\xi)\bar\chi(x;\xi).
\end{split}
\end{equation}
The coefficient functions are members of the scattering data, and $a(\xi)$ can be analytically continued to the upper 
half-plane $\mathop{\rm Im}\xi>0$.
\par
From Eqs.~(\ref{eq:jost_boundaries}) and (\ref{eq:scattering_rels}),
we can see that the Jost function $\phi(x;\xi)$ satisfies the asymptotic form
\begin{equation}
\label{sd}
\phi(x;\xi)=\begin{bmatrix}a(\xi)e^{-i\xi x}\\
b(\xi)e^{i\xi x}\\
\end{bmatrix},\quad\mbox{as $x\to\infty$.}
\end{equation}
When the function $a(\xi)$ has $N$ simple zeros $\xi=\xi_1,\xi_2,\ldots,\xi_N$
on the upper half-plane, $N$ solitons appear in the asymptotic future and each $\xi$ determines the 
characteristics of each soliton. We need to know $a(\xi)$ to extract information on solitons in 
the asymptotic future. By Eq.~(\ref{sd}), we find that this is equivalent to calculating $\phi(x;\xi)$ at $x\to\infty$ 
under the boundary condition given by Eq.~(\ref{eq:jost_neginftyx}).
\section{Discretization of the Initial Wave Packet and Approximated Scattering Data}
In this section, we consider the ZS problem of the NLS equation:
\begin{equation}\label{eq:ZS}
\Psi_x=M\Psi,\quad
M=\begin{bmatrix}
-i\xi&i\psi^\ast\\i\psi&i\xi
\end{bmatrix}.
\end{equation}
Since the spectral parameter $\xi$ is a time-independent quantity, we can take $\psi$ in Eq.~(\ref{eq:ZS}) to be 
the initial value of the unknown wave packet $\psi(x,0)$. 
\par The major difficulty in analyzing Eq.~(\ref{eq:ZS}) for general initial conditions comes from the fact that $\psi
(x,0)$ depends on the coordinate $x$. In order to overcome this difficulty, according to Boffetta and Osborne's 
idea\cite{ref:Boffetta}, we split the support of $\psi(x,0)$ into many small intervals:
\begin{equation}
I_j:\ x_j\le x<x_{j+1}\quad (j=1,\ldots,N),
\end{equation}
and approximate $\psi(x,0)$ so that it takes a constant value in each interval. We introduce a set of functions $
\psi_j$:
\begin{equation}
\psi_j(x)=
\begin{cases}
V_j&x\in I_j,\\
0&x\notin I_j.  \end{cases}
\end{equation}
The initial value $\psi(x,0)$ is now approximated as
\begin{align}
\psi(x,0)&\simeq\sum_{j=1}^N\psi_j(x),\\
&=
\begin{cases}
V_j&(x\in I_j,\ j=1,2,\ldots,N),\\
0&\mbox{(otherwise).}
\end{cases}
\end{align}
Assuming $\psi(x,0)$ belongs to the class of rapidly decreasing functions, 
we can approximately consider that $\psi(x,0)$ has a compact support.
Within each interval, Eq.~(\ref{eq:ZS}) reads as
\begin{equation}\label{eq:splitted_lax}
\Psi_x=M_j\Psi,\quad
M_j=\begin{bmatrix}
-i\xi&iV^\ast_j\\iV_j&i\xi
\end{bmatrix}.
\end{equation}
We can solve Eq.~(\ref{eq:splitted_lax}) for $x$
satisfying $x\in I_j$ as
\begin{subequations}
\begin{equation}
\Psi(x)=T(X)\Psi(x_j),\quad{}T(X)=\exp(XM_j),
\end{equation}
where $X\equiv x-x_j$ and the matrix $T(X)$ is explicitly written as
\begin{align}
\label{eq:transfer_matrix_unit}
T(X)&=
\begin{bmatrix}
\cos KX-i(\xi/K)\sin KX&i(V^\ast_j/K)\sin KX\\
i(V_j/K)\sin KX&\cos KX+i(\xi/K)\sin KX
\end{bmatrix},\\
\label{eq:transfer_matrix_paramK}
K&=\sqrt{\xi^2+|V_j|^2}.
\end{align}
\end{subequations}
We denote the width of the $j$th bin as
\begin{equation}
x_{j+1}-x_j=L_j,
\end{equation}
and we can see that the Jost function satisfies the relation
\begin{align}
\label{eq:transfer_matrix}
&\Psi(x_{N+1})=T\Psi(x_1),\\
&T=T(L_N)T(L_{N-1})\cdots T(L_2)T(L_1).
\end{align}
The matrix $T$ is interpreted as a transfer matrix that connects two asymptotic forms in $x\to\pm\infty$. 
Recalling Eq.~(\ref{eq:jost_neginftyx}) and the fact that we truncate $\psi(x;\xi)$ so that it is supported only in the region 
$x_1\le x\le x_{N+1}$, one can derive the relation
\begin{align}
\phi(x_{N+1};\xi)&=
T\phi(x_1;\xi)
=e^{-i\xi x_1}T\begin{bmatrix}1\\0\end{bmatrix}\\
&=
\begin{bmatrix}a(\xi)e^{-i\xi x_{N+1}}\\
b(\xi)e^{i\xi x_{N+1}}\end{bmatrix}.
\end{align}
Thus, we have the approximated expressions for scattering data in terms of the transfer matrix as
\begin{subequations} 
\label{eq:scattering_data}
\begin{align}
\label{eq:scattering_dataA}
a(\xi)&=e^{iL\xi}T_{11},\\
\label{eq:scattering_dataB}
b(\xi)&=e^{-i(x_1+x_{N+1})\xi}T_{21},
\end{align}
\end{subequations}
where the parameter $L$ is defined as $L=L_1+L_2+\cdots+L_N$.
By considering the initial packet as a set of constant functions, one can obtain explicit expressions for the 
scattering data $a(\xi)$ and $b(\xi)$ for any initial values provided they belong to the class of rapidly decreasing 
functions. Thus, the desired information that characterizes solitons in the asymptotic future can be extracted from $ 
a(\xi)$ with arbitrary precision by suitably adjusting the width of each bin $L_j$.
\section{Applications to Box-Type Initial Conditions}
In this section, we apply Boffetta and Osborne's method introduced in the previous section to box-type initial 
wave packets. We can solve the ZS problems for these initial conditions exactly, and can derive the corresponding final states explicitly.
\par
Throughout this section, we assume that the initial conditions are real-valued, which means that the initial wave packets 
are static. Thus, the spectral parameter $\xi$ is expected to be purely imaginary. Since the zeros of $a(\xi)$ should 
be located in the upper half-plane of $\xi$, we find the discrete eigenvalues under the condition
\begin{equation}
\label{eq:imaginary_specpara}
\xi=i\eta,\quad(\eta>0).
\end{equation}
\subsection{Single-box-type initial condition}
We consider a box-type initial condition whose width is $L$:
\begin{equation}
 \psi(x,0)=
\begin{cases}
V_0&(0\le x\le L),\\
0&(\mbox{otherwise}),
\end{cases}
\end{equation}
where $V_0$ is a real number. From Eqs.~(\ref{eq:transfer_matrix_unit}) and (\ref{eq:scattering_dataA}),
the scattering datum $a(\xi)$ is derived as
\begin{equation}
\label{eq:scatdata_singlebox}
\begin{split}
a(\xi)&=e^{i\xi L}(\cos KL -i\dfrac\xi{K}\sin KL),\\
K&=\sqrt{V_0^2+\xi^2}.
\end{split}
\end{equation}
Under the condition (\ref{eq:imaginary_specpara}), we find that the zeros of $a(\xi)$ can be
derived from the set of relations
\begin{subequations}
\label{eq:zeroconditions}
\begin{align}
\label{eq:zerocondition1}
\sqrt{V_0^2-\eta^2}&=-\eta\tan(L\sqrt{V_0^2-\eta^2}),&(|V_0|>\eta),\\
\label{eq:zerocondition2}
\sqrt{\eta^2-V_0^2}&=-\eta\tanh(L\sqrt{\eta^2-V_0^2}),&(|V_0|<\eta).
\end{align}
\end{subequations}
Since the solution of Eq.~(\ref{eq:zerocondition2}) does not satisfy the condition $\eta>0$, we eliminate it 
and consider only Eq.~(\ref{eq:zerocondition1}).
Introducing $A$ and $u$ as
\begin{equation}
\label{eq:normalized_amp_and_specpara}
A=V_0L,\quad u=\eta L,
\end{equation}
we can omit the parameter $L$. Thus, the equation we should consider becomes
\begin{subequations}
\label{eq:eigeneqs}
\begin{align}
\label{eq:eigeneq_rel}
\sqrt{A^2-u^2}&=-u\tan(\sqrt{A^2-u^2}),\\
\label{eq:eigeneq_cond}
0&<u<A.
\end{align}
\end{subequations}
This means that we should find the intersection of the curves
$y=-u\tan\sqrt{A^2-u^2}$ and $y=\sqrt{A^2-u^2}$ in the first quadrant.
\par
\begin{figure}
\begin{center}
\includegraphics[width=86mm]{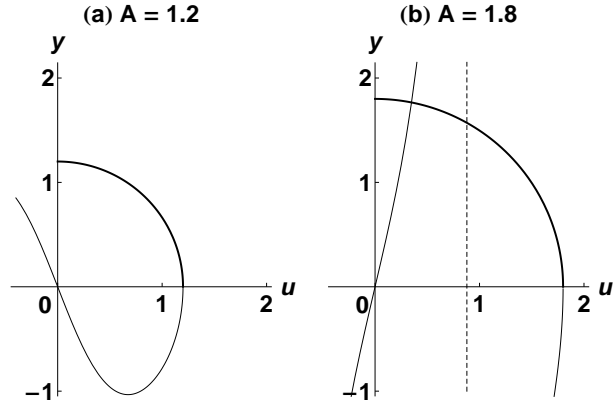}
\end{center}
\caption{Typical cases concerning Eq.~(\ref{eq:eigeneqs}).
(a) Case of $A=1.2$. (b) Case of $A=1.8$.
The thick curves are the graphs of $y=\sqrt{A^2-u^2}$ in the region $u>0$, and the thin curves are the graphs of $y=-u\tan
\sqrt{A^2-u^2}$. The dashed line in (b) denotes the value of $u$ where $\sqrt{A^2-u^2}=\pi/2$ and the right-hand 
side of Eq.~(\ref{eq:eigeneq_rel}) diverges.}
\label{fig:sec4-1}
\end{figure}

For a sufficiently minute value of $V_0$, the value of $\sqrt{A^2-u^2}$ remains in the interval $(0,\pi/2)$ and the 
right-hand side of (\ref{eq:eigeneq_rel}) remains negative. In such a case, no solutions exist and no solitons remain at $t\to\infty$. As the value of $V_0$ increases, $\sqrt{A^2-u^2}$ can exceed $\pi/2$ and solutions of 
Eq.~(\ref{eq:eigeneq_rel}) appear. It is clear that the condition under which Eq.~(\ref{eq:eigeneqs}) has at least one solution is $A>\pi/2$ ($V_0>\pi/(2L)$). Typical situations in which these cases occur are shown in Fig.~\ref{fig:sec4-1}.
After a brief consideration, we can see that if the potential height $V_{0}$ satisfies
\begin{equation}
(n-\dfrac12)\dfrac\pi{L}<V_0\le(n+\dfrac12)\dfrac\pi{L},\quad
\mbox{($n$: a positive integer)},\label{eq:cond}
\end{equation}
the number of solitons that will remain over time should be $n$.
\subsection{Double-box-type initial condition}
\begin{figure}
\begin{center}
\includegraphics[width=60mm]{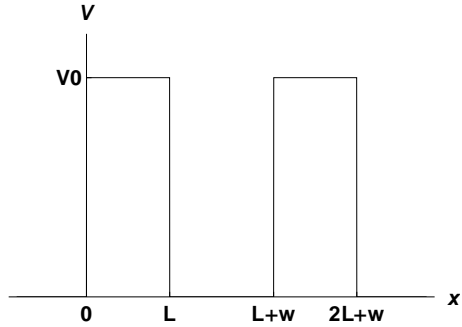}
\end{center}
\caption{Double-box-type initial condition (\ref{eq:doublebox_ic}).}
\label{fig:sec4-2}
\end{figure}

Next, we consider the initial condition
\begin{equation}
\label{eq:doublebox_ic}
\psi(x,0)=
\begin{cases}
V_0,&(0<x<L,\ L+w<x<2L+w),\\ 
0,&(\mbox{otherwise}),
\end{cases}
\end{equation}
which means that two identical pulses, each of which has a common amplitude
$V_0$ and width $L$, are located with separation $w$ (Fig.~\ref{fig:sec4-2}).
In this case, by using Eq.~(\ref{eq:imaginary_specpara}), we find
that the scattering amplitude $a(\xi)$ is given by
\begin{subequations}
\label{eq:scatdata_doublebox}
\begin{align}
a(i\eta)&=e^{-2\eta L}
\Bigg\{\left[\cos KL+\dfrac{\eta}{K}\sin KL\right]^2\nonumber\\
\label{eq:scatdata_A_doublebox}
&\hspace{25mm}-\dfrac{V_0^2}{K^2}e^{-2\eta w}\sin^2KL\Bigg\},\\
\label{eq:scatdata_param_doublebox}
K&\equiv\sqrt{V_0^2-\eta^2}.
\end{align} 
\end{subequations}
If we introduce $A$ and $u$ as in Eq.~(\ref{eq:normalized_amp_and_specpara}),
the zeros of $a(i\eta)$ can be derived from
\begin{subequations}
\begin{align}
\label{eq:doublebox_zerocond_rel}
&\left[\cos K+\dfrac{u}K\sin K\right]^2=
\dfrac{A^2}{K^2}e^{-2uw/L}\sin^2K,\\
&K=\sqrt{A^2-u^2}.
\end{align}
The values of $u$ are restricted to 
\begin{equation}
\label{eq:doublebox_zerocond_reg}
0<u<A,
\end{equation}
\end{subequations}
because there are no positive $\eta$ satisfying this relation if $\eta>V_0$, following the discussion in deriving 
Eq.~(\ref{eq:eigeneq_cond}).
\par
Let us consider two limiting cases. When two initial pulses are sufficiently separated, it is natural to expect that the 
number of solitons that remain over time will be twice that in the case of a single initial pulse, because the amplitude of diffusing 
radiation is generally so small that the interaction between the two pulses hardly affects the asymptotic future. This 
observation is confirmed by taking the limit $w\to\infty$ in Eq.~(\ref{eq:scatdata_A_doublebox}). This operation 
makes the final term in Eq.~(\ref{eq:scatdata_A_doublebox}) vanish. In this limit, the function $a$ given by 
Eq.~(\ref{eq:scatdata_A_doublebox}) coincides with the square of the scattering amplitude of Eq.~(\ref
{eq:scatdata_singlebox})
under the condition (\ref{eq:imaginary_specpara}). In the opposite limit, $w\to0$, the two initial pulses are fused 
together into a single pulse whose width is $2L$. In fact, Eq.~(\ref{eq:scatdata_A_doublebox}) coincides with Eq.~
(\ref{eq:scatdata_singlebox}) if $L$ is replaced by $2L$.
\par
For an appropriately chosen value of $w$, the analysis of the eigenvalue problem provides nontrivial solutions where 
the final term of Eq.~(\ref{eq:scatdata_A_doublebox}) plays an essential role.
\begin{figure}
\begin{center}
\includegraphics[width=86mm]{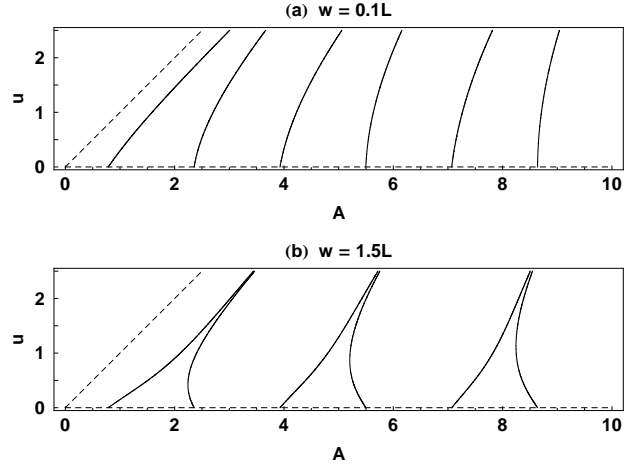}
\end{center}
\caption{Distribution of the zeros of scattering datum
$a(i\eta)$ for various values of $A$ under the given $w$.
(a) Case of $w=0.1L$. (b) Case of $w=1.5L$.
The dashed line expresses the two boundaries of the allowed region for the solutions, given by Eq.~(\ref
{eq:doublebox_zerocond_reg}).
}
\label{fig:sec4-3}
\end{figure}
\begin{figure}
\begin{center}
\includegraphics[width=70mm]{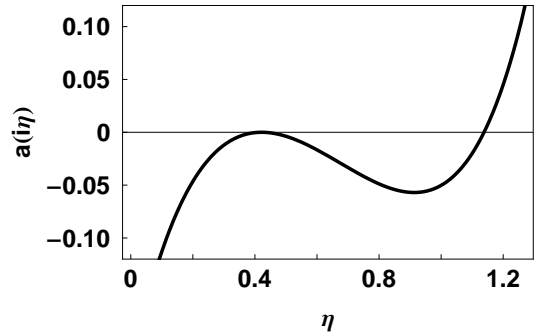}
\end{center}
\caption{Graph of $a(iu)$ for $A=A_0$ ($2.24228$). A double-pole solution of $a(iu)=0$ appears.}
\label{fig:sec4-4}
\end{figure}
We have shown in Fig.~\ref{fig:sec4-3} the curves that satisfy Eq.~(\ref{eq:doublebox_zerocond_rel}) on the $A$-$u$ plane. 
We have chosen the values of separation as
$w=0.1L$ in Fig.~\ref{fig:sec4-3}(a) and $w=1.5L$ in Fig.~\ref{fig:sec4-3}(b). In Fig.~\ref{fig:sec4-3}(a), we can see that there is 
no solution for $A\lesssim0.8$. This means that an initial wave packet with a very small amplitude is completely transformed into 
diffusing waves known as radiation. As the value of $A$$(\sim V_{0})$ increases, a solution of Eq.~(\ref
{eq:doublebox_zerocond_rel}) appears. This occurs when $\pi/4\lesssim A\lesssim 3\pi/4$. 
This solution gives one asymptotically remaining soliton. The soliton is expected to be located midway between the two initial pulses because of the spatial reflection symmetry. For larger values of $A$, the number of 
remaining solitons increases monotonically.
\par
In Fig.~\ref{fig:sec4-3}(b), we can see that there is a qualitatively different result that cannot be observed in 
the previous case. In this case, the quantity $\eta$ is not always given as a single-valued function of $A$ on 
every branch. After having one solution for $\pi/4\lesssim A\lesssim2.2$, Eq.~(\ref
{eq:doublebox_zerocond_rel}) is observed to have two roots around $A\sim2.2$. The smaller solution of $u$ for this 
value of $A$ gives a double-pole solution. We have presented a graph of $a(i\eta)$ in Fig.~\ref{fig:sec4-4} for the value of $A\simeq2.2$, the smallest value of $A$ where the tangent of the curves shown in Fig.~\ref
{fig:sec4-3}(b) is parallel to the $u$-axis.
Let us denote this value of $A$ as $A_0$. When $A=A_0$, the number of remaining solitons is two. If $A$ 
exceeds this value, the number of solitons first becomes three. As the value of $A$ becomes 
larger, the number of remaining solitons decreases to two for $A\gtrsim 3\pi/4$. Thus, the number of 
solitons that appears over time is not a simple monotonic function of the amplitude of the initial 
pulse for a moderate value of $w$.
\subsection{Asymmetric double-box-type initial condition}
In a real experiment on a self-focusing BEC system\cite{ref:Nguyen}, Nguyen {\it et al}. prepared two 
condensates so that the population of one condensate was about half that of the other condensate and made them collide with each 
other. To give the two condensates opposite initial velocities, a harmonic potential was applied in the axial 
direction and they did not turn off the harmonic potential throughout the runs. Unfortunately, the harmonic 
potential was not so weak that its effect could be neglected because the two condensates were observed to pass 
through each other and oscillate back and forth at the bottom of the trap for several periods.  
\par
\begin{figure}
\begin{center}
\includegraphics[width=86mm]{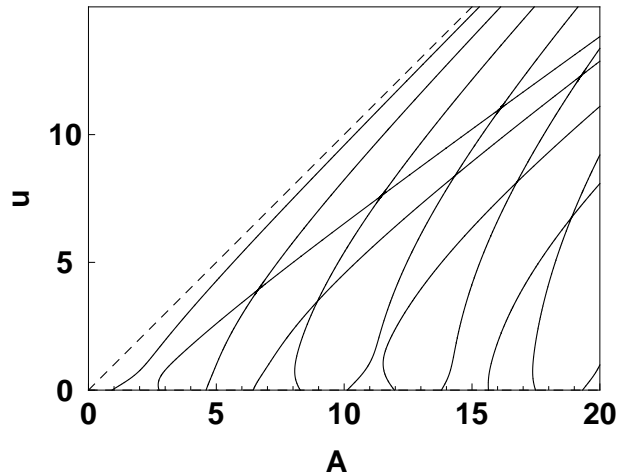}
\end{center}
\caption{Distribution of the zeros of scattering amplitude
$a(iu)$ for various values of $A$ under $w=1.5L$ and $s=1/\sqrt{2}$.
The dashed line expresses the two boundaries of the allowed region for the solutions, given by Eq.~(\ref
{eq:doublebox_zerocond_reg}).
}
\label{fig:sec4-5}
\end{figure}

Nevertheless, from the standpoint of nonlinear wave theory, it is very interesting to consider this 
problem under the ideal potential-free situation. Although the exact initial conditions include smooth shapes and phases of the condensates, we modeled the two condensates as an asymmetric double box-type initial 
condition on the flat line for simplicity as below: 
\begin{equation}
\label{eq:asymdoublebox_ic}
\psi(x,0)=
\begin{cases}
V_0,&(0<x<L),\\
sV_{0},&(L+w<x<2L+w),\\ 
0,&(\mbox{otherwise}).
\end{cases}
\end{equation} 
Assuming $\xi=i\eta$ $(\eta>0)$ and using the normalized parameters $A=LV_{0}$ and $u=L\eta$ as before, we define the
following functions:
\begin{subequations}
\label{eq:scatdata_doublebox_asymmetric1}
\begin{align}
f(A,u)=\cos K(A)+\dfrac{u}{K(A)}\sin K(A),\\
g(A,u)=\frac{A}{K(A)}\sin K(A),\\
K(A)=\sqrt{A^2-u^2}.
\end{align} 
\end{subequations}
The scattering amplitude $a(i\eta)$ for the initial condition (\ref{eq:asymdoublebox_ic}) is now expressed as
\begin{equation}
a(iu)=e^{-2u}\left(f(A,u)f(sA,u)-e^{-2\frac{uw}{L}}g(A,u)g(sA,u)\right),
\end{equation}
which is a symmetric function of $V_{0}$ and $sV_{0}$ as expected, and the last term exactly describes the effect of 
interacting tails. This time, we should note that $f(sA,u)$ and $g(sA,u)$ include hyperbolic functions when $sA<u$. 
In Fig.~\ref{fig:sec4-5}, we show the curves $a(iu)=0$ on the $A$-$u$ plane obtained by setting $w=1.5L$ as before and assuming $s=1/\sqrt{2}$. This selection of $s$ corresponds to the situation where the population of the condensate with a small amplitude is half that of the other condensate, as in the experiment described in Ref. 29.
\begin{figure}
\begin{center}
\includegraphics[width=70mm]{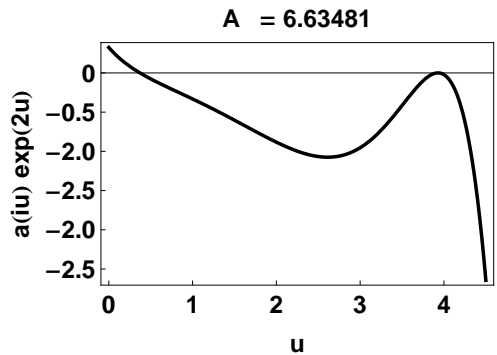}
\end{center}
\caption{Graph of $a(iu)\mathrm{exp}(2u)$ for $A=6.63481$.
A double-pole solution also appears at the crossing point.}
\label{fig:sec4-6}
\end{figure}
As $A$ $({\sim}V_{0})$ increases, we can see, in addition to the fact that $\eta$ is 
not always given as a single-valued function of $A$ on every branch, that some of the branches intersect each other. The first intersection appears at around $A=6.63481$. This point gives a double-pole soliton condition, 
although the measure of this point is no more than zero. We have presented a graph of $a(iu)$ in Fig.~\ref{fig:sec4-6} for this value of $A$.

The appearance of these intersections clearly makes the trajectory $a(iu)=0$ more complex and richer than that in the 
symmetric case. The number of solitons that appears over time is again not a simple monotonic function 
of the amplitude of the initial pulse for a suitably chosen value of $w$.

\section{Numerical Simulation}
We show the results of the numerical simulation of the initial value problem analyzed in the 
previous subsection. By numerically integrating the NLSE (\ref{NLSE}), we solve the initial value problem 
under the double-box-type initial condition (\ref{eq:doublebox_ic}) horizontally shifted so that the center of 
the valley coincides with the origin $x=0$. We set the width of the valley $w$ to $1.5L$ and vary the common 
potential height $V_{0}$  for each time. 

\begin{figure}
\begin{center}
\includegraphics[width=80mm]{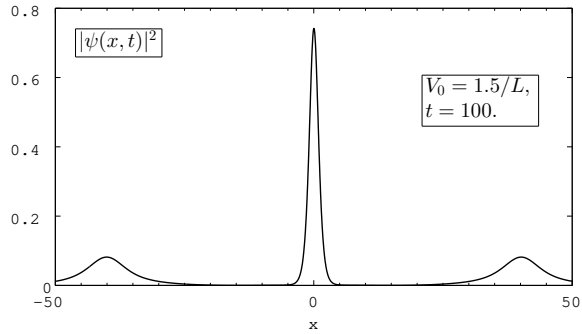}
\end{center}
\caption{Graph of $|\psi(x,t)|^2$ at $t=100$ for $V_{0}=1.5/L$.}
\label{fig:1.5}
\end{figure}

First, we examine the case where $V_{0}=1.5/L$. In this case, we have only one solution so that we expect only one 
soliton in the final state. Figure \ref{fig:1.5} shows the absolute square of the wave amplitude, $|\psi(x,t)|^2$, at 
$t=100$. Although we can observe three pulses, the height of the two smaller peaks relative to that of the center peak keeps decreasing and fades away. Only one soliton at the center is expected to remain at the limit $t\to\infty$.

\begin{figure}
\begin{center}
\includegraphics[width=80mm]{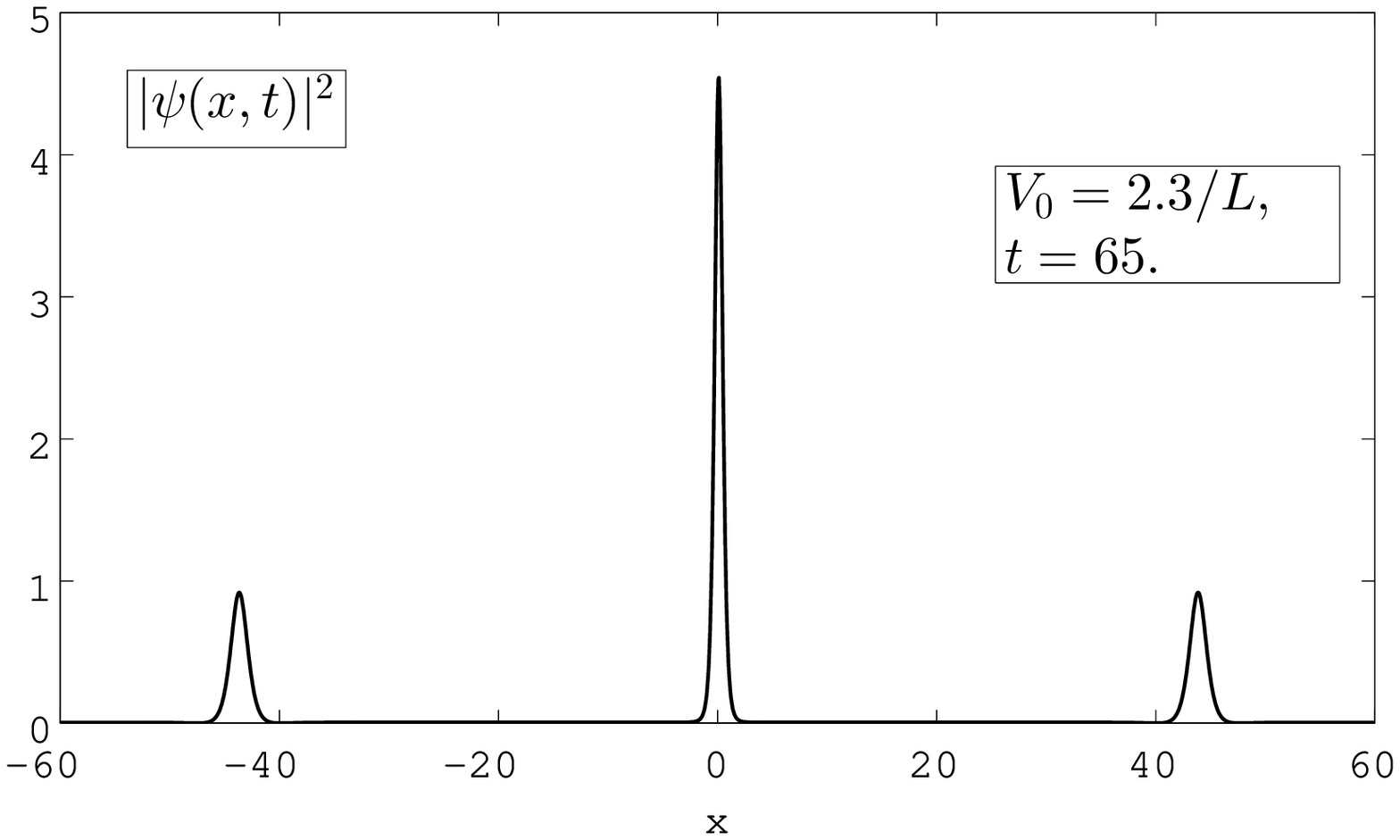}
\end{center}
\caption{Graph of $|\psi(x,t)|^2$ at $t=65$ for $V_{0}=2.3/L$.}
\label{fig:2.3}
\end{figure}

Secondly, we increase the height of the potential to $V_{0}=2.3/L$, which is slightly larger than the critical value 
for a double-pole soliton but smaller than the upper threshold $V_{0}=3\pi/4$. Therefore, we expect three 
remaining solitons in the far distant future. Figure \ref{fig:2.3} shows the value of $|\psi(x,t)|^2$ at $t=65$. We can observe three sharp pulses. In this case, the two smaller peaks at both sides do not vanish.

\begin{figure}
\begin{center}
\includegraphics[width=80mm]{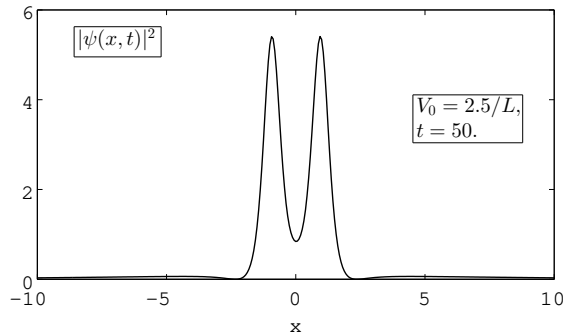}
\end{center}
\caption{Graph of $|\psi(x,t)|^2$ at $t=50$ for $V_{0}=2.5/L$.}
\label{fig:2.5}
\end{figure}

Thirdly, we set the potential height to $V_{0}=2.5/L$, which exceeds the boundary value of $V_{0}=3\pi/4$, and two solitons are predicted to survive. Figure \ref{fig:2.5} shows the absolute square of the wave amplitude 
$|\psi(x,t)|^2$ at $t=50$. We can observe two large peaks around the origin as expected. These peaks keep alternately 
splitting and fusing together, like a breather solution.

In the case of a symmetric initial condition, we also observed similar behaviors. Unfortunately, our numerical 
simulation could not capture the features of the double-pole solitons. This is reasonable because the conditions for generating 
them have zero measures. Other results of the numerical simulation, however, show good agreement with theoretical 
predictions and strongly support their validity.
\section{Discussion and Concluding Remarks}
Inspired by a recent collision experiment using BEC pulses, we have applied Boffetta and Osborne's approximation 
method to analyze the Zakharov-Shabat eigenvalue problem, which is associated with the NLSE. As we have seen in this paper, their method can be used effectively in solving the ZS eigenvalue equations under general nonsoliton initial conditions. We have considered the initial value problem of the self-focusing-type NLS equation under box-type initial conditions, for which we can analytically obtain exact results. We have found that the interplay 
between the decaying tails from the initial pulses can affect the asymptotic behaviors, and we succeeded in making 
qualitative predictions including the number of remaining solitons and conditions under which the initial wave 
becomes double-pole solitons. In particular, under an asymmetric initial condition where the boxes have different 
heights, further complex and interesting behaviors have been observed.
\par
Needless to say, the double-box-type initial conditions that we used are idealized mathematical models and we omitted the external trap potential in the axial direction. Our results, however, are not so far from actual experimental conditions. In the first place, it seems possible to turn off or loosen the axial potential immediately after creating two BEC pulses, so that they can freely interfere with each other. In fact, Nguyen {\it et al.} confirmed that their bright BEC soliton did not diffuse and stayed still in an almost flat axial geometry\cite{ref: Nguyen}. The experimental realization of a BEC bright soliton means that, from Eq.~(\ref{eq:cond}), the key parameter $A=LV_{0}$ satisfies $\pi/2<A<3\pi/2$. We have shown that double pulses with moderate separation develop into a three-soliton state if each pulse satisfies $A\sim2.25$. As we have already mentioned, at least $A\geq\pi/2=1.57$ is already feasible. To achieve $A\sim2.25$ without altering the pulse width $L$, we should have an approximately 1.43-fold larger value of $V_{0}$, which is roughly equivalent to double the total number of atoms. To avoid instability or self-collapse of attractively interacting BEC systems, one must keep the total atom number less than the critical value of $N_{c}=0.67a_{r}/|a_{s}|$\cite{ref:Gamal}, where $a_{r}$ and $a_{s}$ are the radial trap size and the s-wave scattering length, respectively. If we can double the radial trap size $a_{r}$, we can satisfy $A\sim2.25$. According to Ref. 29, we can calculate the value of $a_{r}$ and obtain $a_{r}\sim6\mathrm{\mu m}$. In the observation of the first bright soliton trains of the same ${}^{7}\mathrm{Li}$ atoms, their radial trap size was reported to be about $47 \mathrm{\mu m}$\cite{ref:Strecker}. Hence, $a_{r}= 12 \mathrm{\mu m}$ seems to be an experimentally sound parameter.

Finally, we refer to the possible extensions of this work. Although we have limited ourselves to 
the consideration of the NLS equation, this method can be applied to various soliton equations that belong to the 
AKNS system. In addition, we can expect more extensions to integrable equations that belong to other systems, 
such as the Kaup-Newell\cite{ref:KN}
and Wadati-Konno-Ichikawa\cite{ref:WKI} systems. These extensions should be considered as future works, and more 
interesting physics brought about by nonlinear wave interaction is expected to be extracted in an analytically accessible 
manner. 
\begin{acknowledgments}
The authors express their sincere gratitude to Professor Ralph Willox of the University of Tokyo and Professor Ken-ichi Maruno of Waseda University for their interest in this work and stimulating discussions. We also thank Dr. Sander Wahls at Delft University of Technology, who pointed out important prior works on numerical methods. One of the authors (H.~F.) thanks Utsunomiya University for offering opportunities for fruitful discussions.
\end{acknowledgments}


\end{document}